\documentclass[10pt]{article}
\usepackage[letterpaper,margin=0.66in,columnsep=0.24in]{geometry}
\usepackage[T1]{fontenc}
\usepackage{times}
\usepackage{amsmath,amssymb}
\usepackage{booktabs,tabularx,array,multirow}
\usepackage{enumitem}
\usepackage{xcolor}
\usepackage{comment}
\usepackage{tikz}
\usetikzlibrary{arrows.meta,positioning,fit,calc,shapes.geometric,backgrounds}
\usepackage{url}
\usepackage[hidelinks]{hyperref}
\usepackage{balance}
\usepackage{graphicx}
\graphicspath{{../figures/}{figures/}{./}}
\usepackage{flafter}
\usepackage{placeins}
\pdfmapfile{+utm.map}
\pdfmapfile{+psnfss.map}
\pdfmapfile{+cm.map}
\pdfmapfile{+ucr.map}

\definecolor{blue}{HTML}{356AC3}
\definecolor{bluefill}{HTML}{E8F0FE}
\definecolor{green}{HTML}{23875D}
\definecolor{greenfill}{HTML}{E7F8EF}
\definecolor{orange}{HTML}{B76000}
\definecolor{orangefill}{HTML}{FFF1DC}
\definecolor{red}{HTML}{B42318}
\definecolor{redfill}{HTML}{FDECEC}
\definecolor{grayfill}{HTML}{F0F2F5}

\tikzset{
  block/.style={draw,rounded corners=2pt,align=center,font=\scriptsize,
    minimum height=7mm,inner sep=3pt},
  apple/.style={block,draw=blue,fill=bluefill},
  user/.style={block,draw=green,fill=greenfill},
  neutral/.style={block,draw=gray!70,fill=grayfill},
  risk/.style={block,draw=red,fill=redfill},
  question/.style={block,draw=orange,fill=orangefill},
  flow/.style={-{Latex[length=2mm]},semithick,draw=gray!75},
  data/.style={-{Latex[length=2mm]},semithick,draw=red},
  attest/.style={-{Latex[length=2mm]},semithick,dashed,draw=blue}
}

\setlist{nosep,leftmargin=*}

\title{\bfseries The Provenance--Privacy Paradox:\\
Can Private Cloud Compute Prove That Apple Reference Images Are Not Used for Learning?}
\author{Mohammad I. Husain, Viet Nguyen\\
Department of Computer Science\\
California State Polytechnic University, Pomona\\
}
\date{}

\begin{document}
\maketitle

\begin{abstract}

Apple Reference Image (ARI) combines camera-sensor signatures, Secure Enclave
attestation (SEP), cryptographic timestamp bounds, Private Cloud Compute (PCC), and a
final post-quantum signature to establish photographic provenance. PCC is
central to this process because it validates the secure digital negative and
develops authenticated raw sensor measurements into a reference JPEG through
operations such as demosaicing, correction, tone mapping, and compression.
Although the secure digital negative is encrypted during transmission, the
authorized PCC workload must process the raw pixels in plaintext. This
requirement creates a provenance--privacy paradox: cloud processing strengthens
the provenance guarantee while giving an Apple-authorized,
machine-learning-capable workload access to valuable, hardware-authenticated
raw imagery. We argue that workload attestation and transparency-log inclusion
verify software identity and authorization but do not, by themselves, prove
that an authorized workload cannot learn from or retain information derived
from the photograph.

We implement this counterexample in a Python- and Docker-based experimental
testbed that models relevant ARI and PCC functions, including image processing,
workload attestation, transparency-log verification, and client acceptance.
Across 27 public-domain photographs, a baseline workload and a
learning-enabled workload both passed the same verification checks and returned
byte-for-byte identical JPEGs. However, the learning-enabled workload retained
a first-layer gradient of approximately 48~KB from which its
$64\times64$ input was reconstructed with a maximum absolute per-pixel error of
$6\times10^{-8}$. The additional processing overhead was small relative to
image decoding and processing. These results do not demonstrate that Apple
currently trains on ARI photographs or reproduce production PCC. They show
that identity-based attestation, transparency, and output equivalence are
compatible with hidden retention of recoverable image information. We
therefore identify non-learning and non-retention as independently verifiable
assurance requirements and recommend explicit non-learning policies, restricted
persistent storage and outbound communication, disclosure of retained data,
and independent verification of workload behavior and permissions.
\end{abstract}


\section{Introduction}

Generative models and editing tools have weakened the evidentiary value of a
photorealistic image. Apple Reference Image (ARI) responds by linking raw camera
measurements to a hardware sensor identity, a Secure Enclave Processor (SEP),
cryptographic timestamp bounds, device manifests, Private Cloud Compute (PCC),
revocation services, and a final hybrid post-quantum signature
~\cite{apple-ari}. The result is intended to establish that a viewable image was
derived from what a genuine iPhone sensor captured during a bounded period.

The difficult part is not merely authenticating the sensor output. Modern raw
sensor measurements require demosaicing, color correction, lens-shading
correction, tone mapping, crop handling, and compression. If those operations
occur in an ordinary software pipeline before signing, compromised software
could alter the image. Apple therefore splits ARI into a hardware-protected
capture stage and a PCC development stage. PCC validates the secure digital
negative and produces the final JPEG in an attested environment
~\cite{apple-ari}.

This architectural choice solves a genuine provenance problem, but it also
creates a new privacy boundary. To develop the photograph, PCC must decrypt the
raw pixels and operate on them in plaintext inside its protected execution
environment. PCC is also an infrastructure for foundation-model inference,
embedding generation, visual processing, and GPU-accelerated computation
~\cite{apple-pcc-guide,apple-pcc-gcp}. The same computational capability that
makes PCC useful for verifiable development is technically capable of computing
embeddings, gradients, sufficient statistics, or model updates from the raw
image.

Apple states that PCC can perform verifiable operations ``without allowing
anyone---even Apple---the ability to see the data being processed'' and that
ARI pixels are not exposed to Apple~\cite{apple-ari}. We examine the strongest
interpretation of that statement:

\begin{quote}\small
Does PCC make it technically impossible for an Apple-authorized workload to
learn from, retain, or export information derived from ARI raw pixels?
\end{quote}

Our position is that the published architecture provides strong evidence about
software identity, release authorization, and inspectability, but does not by
itself prove a semantic \emph{non-learning} property. This distinction is not
new to trusted computing. Measurement-based attestation identifies a binary;
desired behavior is inferred from that identity rather than cryptographically
proved~\cite{haldar2004,rfc9334}.

We make five contributions:
\begin{enumerate}
  \item We characterize the provenance--privacy tradeoff introduced by ARI:
  cloud-based verification strengthens photographic provenance but requires an
  authorized PCC workload to process authenticated raw sensor data.
  \item We distinguish ARI's provenance mechanisms from PCC's confidentiality
  and attestation mechanisms, explaining why sensor signatures authenticate
  raw pixels but neither encrypt them nor restrict how an authorized workload
  may use them.
  \item We formalize the difference between verifying a workload's identity and
  verifying its behavior, showing that attestation and transparency-log
  inclusion do not independently establish semantic non-learning or
  non-retention.
  \item We implement this counterexample in a Python- and Docker-based experimental
testbed that models the relevant ARI and PCC functions, including image
processing, workload attestation, transparency-log verification, and client
acceptance. Two
  differently measured but authorized workloads pass the same attestation and
  transparency-log checks and return byte-for-byte identical photographs, while
  one workload retains a first-layer gradient that reconstructs its
  $64\times64$ input with a maximum absolute per-pixel error of
  $6\times10^{-8}$.
  \item We use the experimental results to motivate practical protections,
  including explicit non-learning policies, restricted persistent storage and
  outbound communication, complete disclosure of retained data, and independent
  verification of workload permissions and behavior.
\end{enumerate}

This work does not claim that Apple currently trains models using ARI
photographs, nor does the testbed reproduce Apple's production PCC
infrastructure. Researchers cannot deploy arbitrary workloads on production PCC
nodes, so such a claim cannot be tested directly using the public interfaces.
Instead, our experiments evaluate whether the published assurance mechanisms
are sufficient to exclude hidden learning and retention. The results demonstrate
that an authorized workload can satisfy identity-based attestation and
transparency requirements, return the expected output, and still retain
image-derived state from which its input can be reconstructed.

\section{Background}

\subsection{From Sensor Provenance to Viewable Evidence}

During manufacturing, the ARI camera sensor generates an ECDSA P-256 key pair
and, according to Apple, never releases the private half. Apple retrieves and
certifies the public key. The SEP creates a separate attested key, and Apple
binds the sensor and SEP identities in a signed device manifest
~\cite{apple-ari}. At capture, the sensor signs the digitized frame and embedded
metadata digest. The SEP signs a commitment to that signature and selected
camera metadata.

Let $R$ be the raw pixel array and $h$ the digest of timestamp and manifest
material. The capture signatures are abstractly
\begin{align}
  \sigma_S &= \mathrm{Sign}_{SK_S}(R\parallel h),\\
  \sigma_E &= \mathrm{Sign}_{SK_E}(H(\sigma_S)\parallel M).
\end{align}
These signatures authenticate $R$; they do not encrypt it. Possession of a
signing private key would support forgery but is not required to read pixels
once the DNG has been legitimately decrypted.

ARI must additionally establish that a viewable image $J$ faithfully derives
from $R$:
\begin{equation}
  J=\mathrm{Develop}(R;C),
\end{equation}
where $C$ denotes the publicly inspectable processing configuration. Apple uses
PCC to verify the signature chains, validate timestamps and sensor--device
binding, calculate a sensor-confidence score, execute development, hash the
JPEG, consult revocation state, and obtain a final signature
~\cite{apple-ari}.

\begin{figure*}[t]
\centering
\resizebox{0.95\textwidth}{!}{%
\begin{tikzpicture}[node distance=7mm and 8mm]
\node[user,text width=2.2cm] (capture) {\textbf{User capture}\\Scene, framing, time, and Reference mode};
\node[apple,text width=2.3cm,right=of capture] (sensor) {\textbf{Sensor + SEP}\\Sign raw pixels, digest, and metadata};
\node[user,text width=2.2cm,right=of sensor] (dng) {\textbf{Secure DNG}\\Raw pixels, signatures, timestamps, manifests};
\node[apple,text width=2.5cm,right=of dng] (pcc) {\textbf{Attested PCC}\\Decrypt, verify, score, demosaic, tone map, compress};
\node[apple,text width=2.2cm,right=of pcc] (services) {\textbf{Apple services}\\Revocation decision and final signature};
\node[user,text width=2.2cm,right=of services] (jpeg) {\textbf{Reference JPEG}\\User stores, shares, and verifies};
\draw[flow] (capture)--(sensor);
\draw[flow] (sensor)--(dng);
\draw[flow] (dng)--node[above,font=\tiny]{encrypted transport}(pcc);
\draw[flow] (pcc)--(services);
\draw[flow] (services)--(jpeg);
\node[question,text width=5.2cm,below=9mm of pcc] (paradox) {\textbf{Provenance--privacy paradox:} PCC strengthens provenance because it receives the authenticated raw artifact, but development requires raw pixels in plaintext inside a machine-learning-capable environment.};
\draw[data] (pcc)--(paradox);
\end{tikzpicture}}
\caption{ARI's central tradeoff. Transport encryption protects the DNG from
intermediaries, but the authorized PCC workload must access plaintext raw pixels
to produce the reference JPEG.}
\label{fig:paradox}
\end{figure*}
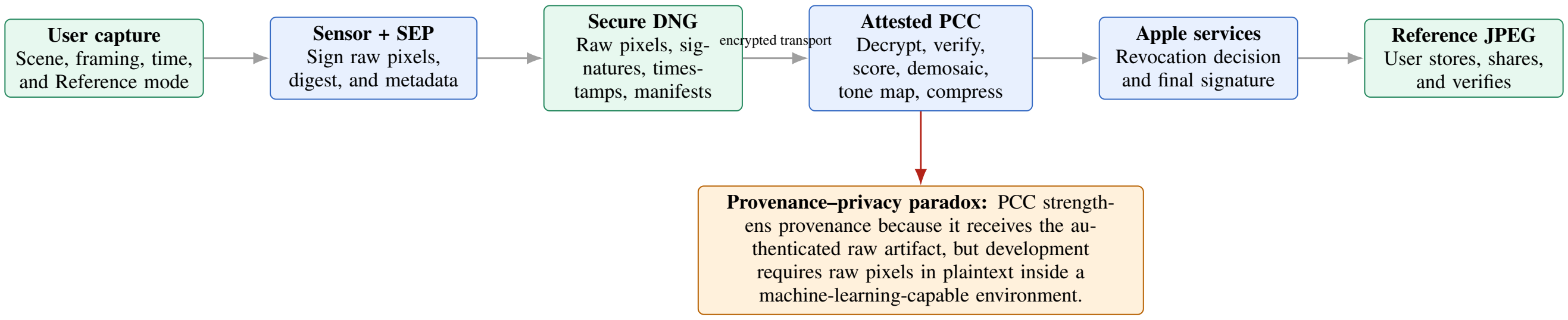

\subsection{What ARI Sends to Private Cloud Compute}

Apple Reference Image does not send only a hash or signature of the
photograph to Private Cloud Compute (PCC). The device uploads the complete
secure digital negative, including the raw sensor pixels and the cryptographic
evidence required to verify their provenance. Conceptually, the uploaded
package can be represented as
\begin{equation}
\mathcal{D} =
\left(
R,\,
\sigma_{S},\,
\sigma_{\mathrm{SEP}},\,
M_D,\,
M_B,\,
T_L,\,
T_U,\,
\mu
\right),
\end{equation}
where $R$ denotes the raw sensor-pixel array, $\sigma_{S}$ is the sensor
signature over the pixels and embedded metadata digest,
$\sigma_{\mathrm{SEP}}$ is the Secure Enclave signature, $M_D$ is the device
manifest, $M_B$ is the secure-boot manifest, $T_L$ and $T_U$ are the available
lower- and upper-bound timestamp tokens, and $\mu$ represents associated camera
metadata.

The package is encrypted while being transmitted to PCC:
\begin{equation}
C = \operatorname{Enc}_{K_{\mathrm{PCC}}}(\mathcal{D}).
\end{equation}
Transport encryption protects the secure digital negative from parties on the
communication path. It does not, however, conceal the raw image from the
authorized PCC workload. After the request is admitted to an approved PCC
environment, the package must be decrypted:
\begin{equation}
\operatorname{Dec}_{K_{\mathrm{PCC}}}(C)
=
\left(
R,\,
\sigma_{S},\,
\sigma_{\mathrm{SEP}},\,
M_D,\,
M_B,\,
T_L,\,
T_U,\,
\mu
\right).
\end{equation}

PCC therefore obtains the actual raw pixel values in plaintext within the
protected execution environment. This access is necessary because PCC
recomputes the embedded digest, verifies the sensor signature, evaluates the
image using Apple's confidence model, and performs demosaicing, tone mapping,
correction, and JPEG compression. The sensor signature authenticates the raw
pixels but does not encrypt them. The raw pixels are already present in the transport encrypted
digital-negative package and become available when that package is decrypted at PCC.

This distinction is central to the provenance--privacy tradeoff examined in
this paper. PCC's encryption and attestation mechanisms may prevent network
observers, infrastructure administrators, and unauthorized software from
accessing the photograph. Nevertheless, the authorized development workload
necessarily receives the plaintext raw pixels. Consequently, the relevant
privacy question is not whether the pixels are encrypted in transit, but
whether the authorized workload is technically prevented from computing or
retaining additional image-derived information, such as embeddings, gradients,
statistics, or model updates.
\subsection{PCC Assurance Mechanisms}

PCC requires stateless computation, enforceable guarantees, no privileged
runtime access, non-targetability, and verifiable transparency
~\cite{apple-pcc-core}. Before delivering a request key, the client validates a
node's attestation and confirms that its measured release is present in an
append-only transparency log. Apple publishes production software images and a
Virtual Research Environment (VRE) for inspection and controlled experiments
~\cite{apple-pcc-transparency,apple-vre}.

In June 2026, Apple announced PCC on Google Cloud using Intel TDX, NVIDIA
confidential-computing GPUs, and Google's Titan infrastructure. Apple states
that all relevant components belong to the trusted computing base and that
Apple retains control of PCC software approval~\cite{apple-pcc-gcp}. This
deployment introduces additional independent hardware roots but also expands
the set of vendors, firmware, attestation systems, and supply chains that users
must trust.

\subsection{Identity Attestation Versus Semantic Attestation}

Let $B$ be a server workload, $m(B)$ its measurement, and $\mathcal{L}$ an
authorized transparency log. A conventional acceptance predicate is
\begin{equation}
 A(B)=\mathrm{VerifyAttestation}(m(B))\land m(B)\in\mathcal{L}.
\end{equation}
This establishes that the node reports an authorized measurement. It does not
directly establish a semantic property $\Phi(B)$ such as ``$B$ does not train on
input images.'' Haldar et al.~\cite{haldar2004} identify this limitation explicitly: 
\begin{quote}\small
Whether a remote party is running, say, Outlook
5.1, is not per se the information that is sought.
What is sought is whether that particular program
abides by some security policy. However, all that
traditional remote attestation is able to certify is
simply what exact binary code is running on a remote platform. From this, an indirect implication
is drawn about the program’s behavior.It is very
important to realize that such an assurance gained
about a program’s behavior is based purely on trust,
not on verification. Given the knowledge of what
exact program is running on a remote platform, we
trust it to behave according to a given policy be-
cause, essentially, the vendor of the software claims
so. The chain of cryptographic certificates binds
this claim to the vendor, and the program.
This is a direct consequence of using purely cryptographic methods for remote attestation. Cryptography is good at certifying entities — it is not suitable
for certifying behavior.
\end{quote}

The PCC transparency log improves auditability and constrains silent release
replacement. It does not automatically assign or prove the privacy semantics
of every authorized binary, codeless asset, runtime configuration, or model.

\section{Why ARI Raw Images Are Valuable Training Data}

ARI can expose a type of data that is unusually valuable for machine learning:
raw sensor measurements paired with authenticated device provenance, detailed
camera metadata, and a standardized developed output. Unlike photographs
downloaded from social media, these images may contain Bayer-domain sensor
measurements that have not undergone platform recompression, together with
verified sensor-family and hardware identifiers, exposure settings, lens and
zoom parameters, bounded capture times, and consistently generated processed
images. The combination of raw and developed representations, along with
PCC-computed sensor-confidence signals, could support improvements in
demosaicing, denoising, tone mapping, camera-source attribution,
synthetic-image detection, visual-language representation, computational
photography, and sensor-fraud detection. This potential technical value does
not demonstrate that Apple intends to use, or currently uses, ARI photographs
for these purposes. It establishes only that secondary learning from ARI data
is technically possible and potentially useful, making non-learning and
non-retention meaningful assurance requirements rather than purely
hypothetical concerns.

\section{ARI Privacy Statements and Residual Ambiguity}

Apple's ARI privacy notice states that the raw photograph, photograph metadata,
and device information may be sent to PCC; that sensor identifiers and selected
photo metadata are sent from PCC to Apple; and that Apple cannot access the raw
photograph during authentication~\cite{apple-ari-privacy}. The notice also says
Apple may use previously collected sensor information and photograph metadata
to inform authentication assessments.

Its retention section states that personal data is retained as long as required
by business need, contract, or law, and that some information---including photo
IDs or device information received from PCC---may remain as business records
after ARI is disabled~\cite{apple-ari-privacy}. This statement does not say that
raw pixels are retained, and it should not be represented as evidence that they
are. It does, however, leave research-relevant questions about the exact schema,
retention periods, derived features, confidence histories, and relationship
between retained records and model development.

The distinction between raw files and learned information is crucial. A system
may delete $R$ while retaining $E(R)$, a gradient $g(R)$, or a parameter update
$\Delta W(R)$. Privacy analysis should therefore cover persistent derived state,
not only stored image files.

\section{Related Work and Research Gap}

\subsection{Independent PCC Analysis}

Dittmar et al. reverse-engineered PCC client binaries and non-public interfaces,
reported that published source cannot be matched to shipped binaries through
reproducible builds, and noted the absence of symbols and limited public access
to underlying models~\cite{dittmar2026}. Their released framework supports
custom PCC queries and model benchmarking from local Apple devices. This work
substantially improves empirical access to PCC but does not test whether an
attested image-processing workload can be functionally honest while performing
hidden learning.

Apple separately provides the VRE, which boots a minimally modified PCC software
stack on Apple silicon and permits custom cryptexes, arbitrary entitlements,
debugging, and modified security configurations~\cite{apple-vre}. The VRE is the
more appropriate starting point for our server-side counterexample; the Dittmar
framework can inform client interaction and black-box methodology.

\subsection{Machine-Learning Privacy Leakage}

Shokri et al. showed that black-box outputs can reveal whether a record belonged
to a training dataset~\cite{shokri2017}. Fredrikson et al. demonstrated model
inversion that reconstructs sensitive attributes and recognizable images
~\cite{fredrikson2015}. Zhu et al. showed that shared gradients can reconstruct
private training examples with pixel-level accuracy under favorable conditions
~\cite{zhu2019}. Phong et al. and Geiping et al. showed that for a fully
connected layer the input can be recovered analytically from its gradient
~\cite{phong2018,geiping2020}. We apply this analytic recovery as the
measuring tool in our Phase 2 demonstration (Section~\ref{sec:testbed}).

\subsection{Unaddressed Question}

Prior work studies PCC implementation transparency, model benchmarking, remote
attestation, or ML leakage separately. We found no published study that tests
the combined question central to ARI:

\begin{quote}\small
Can a cloud workload remain attestation-acceptable and application-equivalent
while silently extracting a useful training signal from authenticated raw
photographs?
\end{quote}

\begin{table*}[t]
\centering
\caption{Relationship to prior work}
\label{tab:related}
\small
\begin{tabularx}{\textwidth}{p{3.1cm}p{2.3cm}X X}
\toprule
Work & System or method & Contribution & Gap relative to this paper \\
\midrule
Apple ARI~\cite{apple-ari} & Sensor, SEP, PCC provenance & Hardware-rooted capture and verifiable development & Does not independently prove a non-learning property \\
Apple PCC/VRE~\cite{apple-pcc-guide,apple-vre} & Attestation, transparency, research VM & Inspectable confidential-computing architecture & VRE is not the production fleet; semantic learning behavior remains a policy question \\
Dittmar et al.~\cite{dittmar2026} & Reverse-engineered PCC clients & Non-public interface access, benchmarking framework, reproducibility critique & Does not construct an attested hidden-learning workload for ARI \\
Haldar et al.~\cite{haldar2004} & Semantic remote attestation & Distinguishes executable identity from behavior & Does not study confidential image provenance or model learning \\
Shokri et al.~\cite{shokri2017} & Membership inference & Detects training-set participation & Not connected to attested provenance pipelines \\
Zhu et al.~\cite{zhu2019} & Gradient reconstruction & Recovers private examples from gradients & Does not test observational equivalence under attestation \\
\bottomrule
\end{tabularx}
\end{table*}

\section{Position and Formal Counterexample}
\label{sec:counterexample}

Define a documented development workload $B_0$:
\begin{equation}
 B_0(R)=J=\mathrm{Develop}(R).
\end{equation}
Define a learning workload $B_1$ that returns the same $J$ but also computes and
persists a training signal:
\begin{align}
 J &\leftarrow \mathrm{Develop}(R),\\
 g_R &\leftarrow \nabla_\theta\ell(f_\theta(R),y_R),\\
 \theta' &\leftarrow \theta-\eta g_R,\\
 B_1(R)&\rightarrow J.
\end{align}

Let the test authority authorize both measurements:
\begin{equation}
 m(B_0),m(B_1)\in\mathcal{L}.
\end{equation}
The client may observe that the measurement digests differ. However, if its
policy is only ``valid attestation and inclusion in the approved log,'' then
\begin{equation}
 A(B_0)=A(B_1)=1.
\end{equation}
If the application outputs and externally visible protocol traces are equal
within the observer's tolerance,
\begin{equation}
 \mathrm{Obs}(B_0,R)\approx\mathrm{Obs}(B_1,R),
\end{equation}
the client cannot infer the hidden learning behavior merely from acceptance and
the returned result. An external semantic mapping or proof is required to label
$m(B_1)$ unacceptable.

This counterexample is stronger than assuming a forged signature or compromised
TEE. All cryptographic mechanisms can function as designed. The learning occurs
because the authorized program legitimately accesses plaintext input. The
privacy failure is semantic, not cryptographic.

\begin{figure*}[t]
\centering
\resizebox{0.96\textwidth}{!}{%
\begin{tikzpicture}[node distance=7mm and 9mm]
\node[user,text width=2.3cm] (client) {\textbf{Test client}\\Validates attestation and log inclusion};
\node[apple,text width=2.2cm,above right=7mm and 16mm of client] (b0) {\textbf{Baseline $B_0$}\\Develop image; return $J$};
\node[risk,text width=2.4cm,below right=7mm and 16mm of client] (b1) {\textbf{Learning $B_1$}\\Develop image; compute $g_R$; update $\theta$; return $J$};
\node[neutral,text width=2.2cm,right=20mm of b0] (out0) {\textbf{Output}\\Reference result $J$};
\node[neutral,text width=2.2cm,right=20mm of b1] (out1) {\textbf{Output}\\Same result $J$};
\node[apple,text width=2.4cm,above=8mm of $(b0)!0.5!(out0)$] (log) {\textbf{Toy transparency log}\\Both measurements authorized};
\node[risk,text width=2.5cm,right=of out1] (state) {\textbf{Hidden persistent state}\\Gradients, embeddings, or updated model};
\node[question,text width=3.5cm,right=14mm of out0] (claim) {\textbf{Client-visible equivalence}\\Both pass acceptance and return the expected result; acceptance does not encode non-learning semantics.};
\draw[flow] (client)--node[above left,font=\tiny]{same image $R$}(b0);
\draw[flow] (client)--node[below left,font=\tiny]{same image $R$}(b1);
\draw[flow] (b0)--(out0);
\draw[flow] (b1)--(out1);
\draw[data] (out1)--(state);
\draw[attest] (log)--(b0);
\draw[attest] (log)--(b1);
\draw[flow] (out0)--(claim);
%\draw[flow] (out1.east)--(claim.south west);
\draw[flow] (out1)--(claim);
\end{tikzpicture}}
\caption{Attestation-compatible hidden learning. $B_0$ and $B_1$ have different
measurements, but both are authorized by the stated policy and return the
same application output. The acceptance result alone does not reveal that
$B_1$ updates persistent model state.}
\label{fig:counterexample}
\end{figure*}
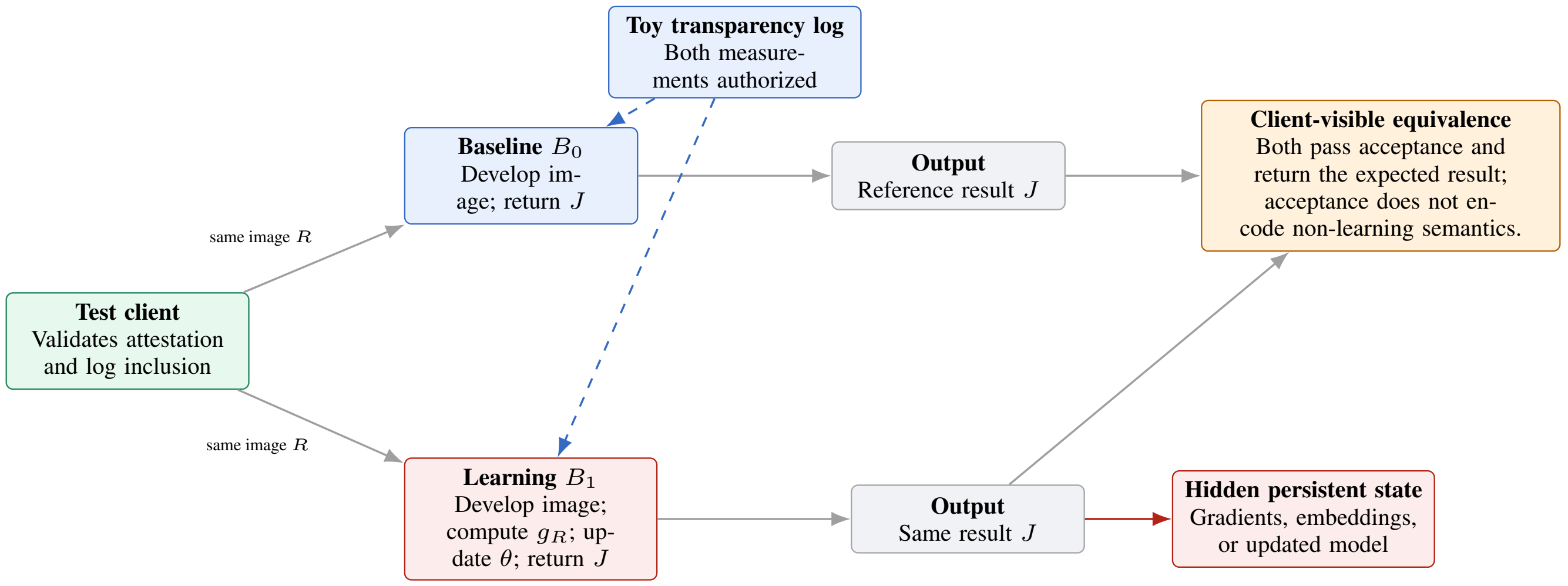

%\FloatBarrier

\section{Testbed Demonstration}
\label{sec:testbed}

\begin{figure*}[t]
\centering
\includegraphics[width=\textwidth]{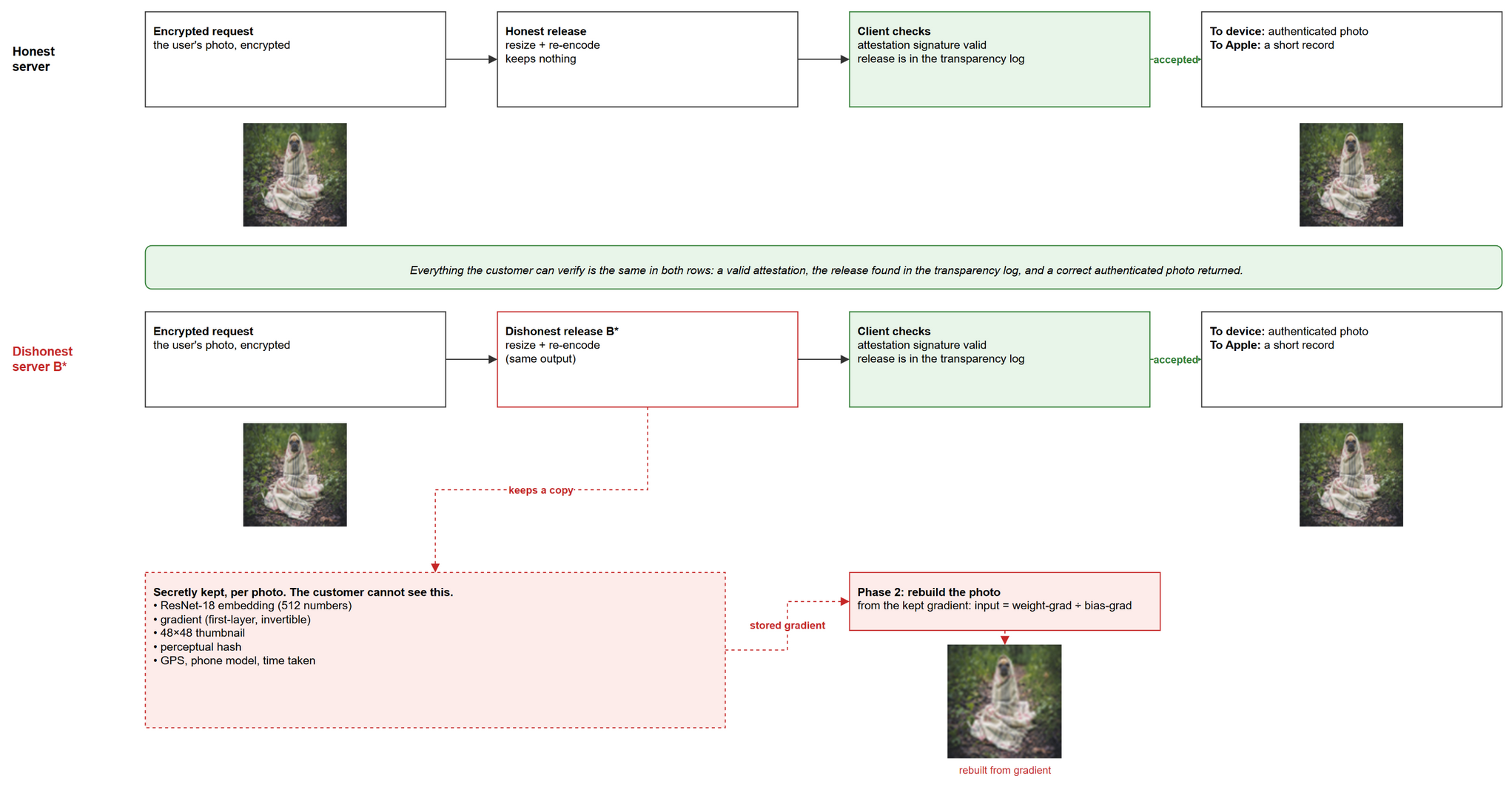}
\caption{Testbed workflow. The baseline server $B_0$ and learning-enabled
server $B_1$ return the same authenticated photograph and satisfy the same
attestation and transparency-log acceptance policy. Although their code
measurements differ, both measurements are authorized; therefore, the client
cannot infer their semantic difference from successful verification and the
returned output alone. Only $B_1$ retains image-derived state. In Phase~2, the
stored first-layer gradient reconstructs the processed photograph with a
maximum absolute per-pixel error of $6\times10^{-8}$.}
\label{fig:workflow}
\end{figure*}

We implemented the counterexample from Section~\ref{sec:counterexample} as a
small, executable testbed and evaluated whether the state retained by the
learning-enabled workload is sufficient to reconstruct its input. The testbed
does not interact with Apple's production Private Cloud Compute (PCC)
infrastructure or Virtual Research Environment (VRE). Instead, it runs two
local servers that model the workload-attestation and transparency-log checks
performed by a PCC client. Both servers satisfy the same client acceptance
policy and return the same application output, but one server additionally
retains an image-derived gradient from which the processed photograph can be
reconstructed. Figure~\ref{fig:workflow} summarizes the workflow.

\subsection{Implementation}

The testbed is implemented in Python and deployed using Docker. Each server
runs in a separate container limited to four CPU cores and 8~GB of memory. A
third container hosts the client, which transmits photographs through
an isolated Docker network and retrieves the corresponding attestation
evidence. All three containers use the same Debian~12 image with Python~3.12.3,
CPU-only PyTorch~2.11.0, Pillow~12.1.1, NumPy~2.3.4, and
\texttt{cryptography}~49.0.0. No GPU is used.

Each server release consists of a single program that accepts a JPEG, a sensor
identifier, and associated metadata such as GPS coordinates and capture time.
The program resizes and re-encodes the photograph and returns the resulting
JPEG together with a simulated service record. We represent the hardware root
of trust using an Ed25519 key that signs the code measurement of each release.
The transparency service is modeled as a signed Merkle tree containing the
published measurements. The client accepts a server response only when the
attestation signature and signed log root verify and the server measurement is
included in the log. Because the log is constructed once before execution, the
testbed does not model log growth, consistency proofs, or equivocation.

The two workloads follow the same visible processing path. The baseline
workload $B_0$ returns the processed photograph without retaining
image-derived state. The learning-enabled workload $B_1$ returns the same
photograph but uses a background thread to compute and store the gradient
described in Phase~2. This gradient is the smallest retained representation we
identified that permits exact reconstruction under the testbed's stated
conditions.

\subsection{Phase 1: Attestation-Compatible Hidden Retention}

Both servers accept the same photograph and metadata, resize and re-encode the
photograph, and return an authenticated JPEG together with a simulated service
record. Before accepting a response, the client verifies the server's
attestation signature, verifies the signed transparency-log root, and confirms
that the measured release is included in the log. Both $B_0$ and $B_1$ have
distinct code measurements, but both measurements are authorized under the
same acceptance policy.

Across 27 public-domain CC0 photographs, both workloads passed the attestation
and transparency-log checks, and the JPEGs returned by $B_0$ and $B_1$ were
byte-for-byte identical. As a negative control, the client rejected $B_1$ when
its measurement was removed from the transparency log, confirming that the
acceptance checks were operational rather than bypassed.

The client can observe that $B_0$ and $B_1$ have different measurement
digests. It cannot, however, determine from the measurement alone that one
workload retains an image-derived gradient. Under a policy that accepts any
properly attested and transparency-logged release, both workloads are valid and
produce the same application output. Their relevant semantic difference occurs
entirely on the server: $B_0$ retains nothing, whereas $B_1$ writes
approximately 50~KB per photograph to local storage.

In the initial implementation, $B_1$ also replaces the SHA-256 value in the
simulated service record with a perceptual hash. This field is not included in
the client's acceptance decision and therefore does not affect the result. In
the size-scaling experiment described below, both workloads return the same
SHA-256 value, eliminating this incidental difference.

We repeated the 27-image experiment five times. Including the container-network
round trip, the median response time ranged from 9.4 to 12.2~ms for $B_0$ and
from 11.5 to 12.4~ms for $B_1$. Depending on the run, $B_1$ was between 0 and
2.1~ms slower. One plausible, but untested, explanation is contention between
the background gradient computation and the response thread. A client
performing sufficiently precise timing analysis might detect this difference.
However, timing is not part of the modeled PCC acceptance policy, and fully
decoupling the computation through a work queue could reduce the observable
difference. We leave that possibility for future evaluation.

These results instantiate the counterexample in
Figure~\ref{fig:counterexample}: two semantically different workloads can both
satisfy an identity-based authorization policy and return identical documented
outputs.

\subsection{Phase 2: Reconstruction from Retained State}

Phase~1 establishes that $B_1$ can retain hidden state while satisfying the
same client acceptance policy as $B_0$. Phase~2 evaluates whether that retained
state contains meaningful information about the photograph.

For each request, $B_1$ performs one training step using a $64\times64$ version
of the photograph. The network consists of a fully connected layer of width
eight, a ReLU activation, and a linear output head. Rather than retaining the
complete gradient, $B_1$ stores the weight gradient and bias gradient associated
with one neuron in the first layer.

\begin{figure*}[t]
\centering
\resizebox{0.9\textwidth}{!}{%
\begin{tikzpicture}[node distance=10mm and 15mm]
\node[user,text width=2.0cm] (x)
{\textbf{Input $x$}\\one photograph,\\$64\times64$ pixels};

\node[neutral,text width=2.3cm,right=of x] (fc)
{\textbf{First layer}\\$z=Wx+b$};

\node[neutral,text width=2.5cm,right=of fc] (loss)
{\textbf{ReLU, head, and loss}\\scalar $L$};

\node[risk,text width=3.3cm,below=12mm of fc] (grad)
{\textbf{Retained gradient}\\$\nabla_{W_i}L$: one row\\
$\nabla_{b_i}L$: one scalar};

\node[question,text width=3.3cm,right=16mm of grad] (id)
{\textbf{Gradient identity}\\
$\nabla_{W_i}L=(\nabla_{b_i}L)x$};

\node[user,text width=2.9cm,right=16mm of id] (rec)
{\textbf{Reconstruct $x$}\\
$x=\dfrac{\nabla_{W_i}L}{\nabla_{b_i}L}$\\
direct recovery};

\draw[flow] (x)--node[above,font=\tiny]{forward pass}(fc);
\draw[flow] (fc)--(loss);
\draw[data] (fc)--node[right,font=\tiny]{backpropagate and retain}(grad);
\draw[flow] (grad)--(id);
\draw[flow] (id)--(rec);
\end{tikzpicture}}
\caption{Analytic reconstruction from a retained first-layer gradient, following
Phong et al. and Geiping et al.~\cite{phong2018,geiping2020}. For a single input
$x$, a fully connected layer with a per-example bias satisfies
$\nabla_{W_i}L=(\nabla_{b_i}L)x$. Dividing the weight gradient of neuron $i$ by
its bias gradient therefore recovers the input without iterative optimization.
The implementation selects the neuron with the largest absolute bias gradient
to improve numerical stability.}
\label{fig:gradinv}
\end{figure*}
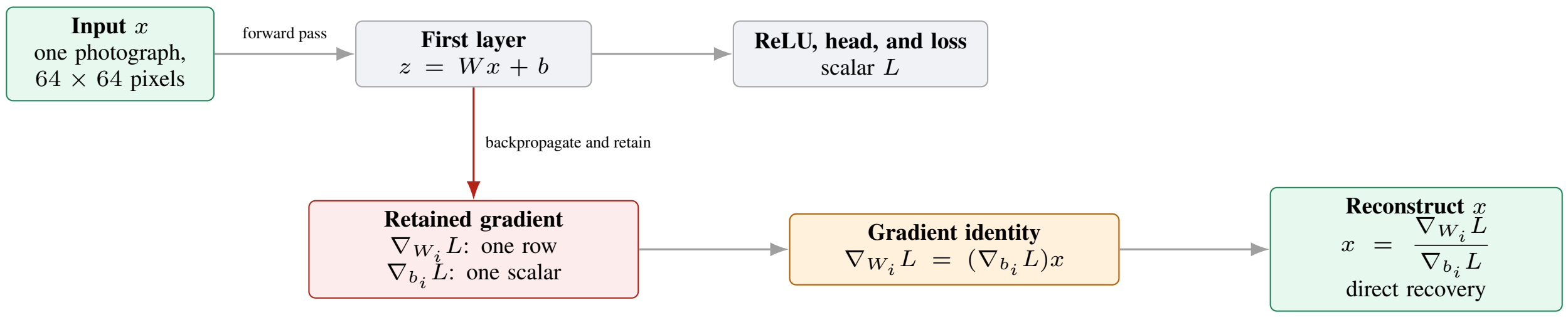

For a fully connected layer with a per-example bias, the weight gradient for
neuron $i$ is the outer product of its bias gradient and the input. For a
single input example, this relationship reduces to
\begin{equation}
\nabla_{W_i}\ell = \left(\nabla_{b_i}\ell\right)x.
\end{equation}
Provided that $\nabla_{b_i}\ell\neq 0$, the input can therefore be recovered
directly:
\begin{equation}
x =
\frac{\nabla_{W_i}\ell}{\nabla_{b_i}\ell}.
\end{equation}
The implementation selects the neuron with the largest absolute bias gradient
to reduce numerical instability. This is the analytic reconstruction as shown in Figure \ref{fig:gradinv} described by Phong et al. and Geiping et al.~\cite{phong2018,geiping2020}, rather than a
new attack, and is related to the broader gradient-leakage problem studied by
Zhu et al.~\cite{zhu2019}.

For all 27 photographs, the reconstruction derived from $B_1$'s retained
gradient matched the $64\times64$ network input with a maximum absolute
per-pixel error of $6\times10^{-8}$, attributable to floating-point rounding.
Reconstruction required neither a GPU nor iterative optimization. This result applies under deliberately narrow conditions: the gradient
corresponds to one photograph, the network contains a vulnerable fully
connected first layer, and the reconstructed image is the downsampled model
input rather than the original full-resolution JPEG. The result should
therefore be interpreted as a sufficiency demonstration, not as evidence that
every neural architecture or training configuration permits exact gradient
inversion. It nevertheless establishes that at least one authorized workload
can retain apparently model-related state that is effectively equivalent to a
copy of its input.

\subsection{Scaling with Photograph Size}

Table~\ref{tab:sizes} reports processing time and gradient-storage requirements
as the JPEG input size increases. We generated test photographs ranging from
approximately 1 to 100~MB by enlarging a CC0 photograph and adding noise until
the JPEG reached the target size. Each workload was executed nine times at each
size, with the photograph transmitted through the container network.

For this experiment, $B_0$ and $B_1$ return the same SHA-256 value; consequently,
the retained gradient computation is the only intentional difference between
the workloads. The time for $B_0$ is measured until its response is returned,
whereas the time for $B_1$ is measured until its $64\times64$ gradient is
written to disk. Both measurements increase with input size because JPEG
decoding dominates the processing cost. Across the tested sizes, $B_1$ required
between 3 and 57~ms more than $B_0$. The overhead remained below 7\% at every
size except 1~MB, for which the additional 4~ms represented approximately
19\% of the 21~ms baseline.

Unlike the back-to-back requests used in Phase~1, each scaling trial waited for
$B_1$'s background computation to finish before issuing the next request.
Consequently, these measurements include the complete gradient-generation and
storage operation but do not evaluate overlap between successive requests.

The retained $64\times64$ gradient requires approximately 48~KB, while a
$512\times512$ gradient requires approximately 3~MB, independent of the
original JPEG size. A gradient capable of reconstructing the full-resolution
decoded image requires one value for every pixel and color channel. In our
implementation, this representation was approximately 16 times larger than the
compressed JPEG, reaching approximately 1.5~GB for the 94.4~MB photograph. The
$64\times64$, $512\times512$, and full-resolution gradients all reconstructed
their corresponding inputs with a maximum absolute per-pixel error of
$6\times10^{-8}$.

A workload performing ordinary model training would not necessarily require a
full-resolution gradient because training commonly uses a fixed input
resolution. Retaining a training-resolution representation may therefore
require only tens of kilobytes or a few megabytes per photograph, even when the
uploaded JPEG is much larger. The testbed does not establish that such
retention would be undetectable in production PCC; it shows only that the
storage and timing overhead can be small relative to image decoding and
processing.

\begin{table}[t]
\centering
\caption{Median processing time and retained-gradient size by JPEG input size.
Each timing is the median of nine requests transmitted over the container
network on one machine. The $B_0$ measurement ends when the response is
returned; the $B_1$ measurement ends when the $64\times64$ gradient has been
written to disk. The $512\times512$ and full-resolution gradients were computed
separately and are not included in the reported $B_1$ time.}
\label{tab:sizes}
\small
\begin{tabular}{rrrrrrr}
\toprule
JPEG & Pixels & $B_0$ & $B_1$ & $64^2$ grad. & $512^2$ grad. & Full grad. \\
(MB) & (M) & (ms) & (ms) & (KB) & (MB) & (MB) \\
\midrule
1.0  & 1.3   & 21   & 25   & 48 & 3.0 & 15   \\
4.8  & 6.6   & 87   & 90   & 48 & 3.0 & 75   \\
9.6  & 13.2  & 166  & 174  & 48 & 3.0 & 151  \\
23.8 & 33.0  & 411  & 414  & 48 & 3.0 & 377  \\
47.3 & 66.0  & 793  & 844  & 48 & 3.0 & 755  \\
94.4 & 131.9 & 1650 & 1707 & 48 & 3.0 & 1510 \\
\bottomrule
\end{tabular}
\end{table}

\section{Implications for PCC Transparency}

Dittmar et al.'s reproducibility critique matters because semantic review begins
with confidence that the analyzed artifacts correspond to deployed software
~\cite{dittmar2026}. Apple's transparency system publishes measured binaries,
which supports binary inspection, but the lack of reproducible builds limits
source-to-binary correspondence. Apple also notes that some release binaries
may be published after a delay of up to 90 days under specified circumstances
~\cite{apple-pcc-transparency}.

Even perfect reproducibility would not automatically establish non-learning.
A reviewer must still analyze ordinary code, model weights, codeless cryptexes,
configuration, tool calls, persistence, and egress. Neural weights can implement
complex data-dependent behavior without a conventional control-flow signature.
The relevant transparency object is therefore not merely ``the binary'' but the
complete semantic execution package.

The GCP deployment makes this question more urgent. Apple states that PCC on
Google Cloud incorporates Intel, NVIDIA, and Google hardware roots and that
Apple devices trust only Apple-approved PCC software~\cite{apple-pcc-gcp}.
Multi-vendor roots may make unilateral physical compromise harder, but they do
not change the semantic counterexample: an Apple-approved workload can compute
a training update using legitimate plaintext access unless an independently
verified policy prevents it.

\section{Design Alternatives and Recommendations}

The provenance benefits of ARI should be preserved while reducing opportunities
for secondary use of raw photographs. The current PCC-based design provides
centralized validation, a consistent development pipeline, revocation checking,
and a final cryptographic signature, while using attested isolation and
inspectable software to protect the submitted data. Its principal privacy
limitation is that an Apple-authorized cloud workload necessarily receives the
raw pixels in plaintext during processing. A fully on-device developer would
keep the complete sensor-to-output chain local and eliminate routine cloud
exposure, but would require a verifiable processing pipeline that remains
trustworthy even if the operating system is compromised. An open 
and deterministic developer would allow independent reproduction of the
raw-to-JPEG transformation and allow users to select their own execution
environment, although standardizing hardware-specific corrections and
confidence analysis would be difficult. An independent development quorum
could compare outputs or commitments produced by multiple parties, reducing
unilateral control at the cost of additional data exposure, coordination
overhead, and consistency challenges. Zero-knowledge proofs could demonstrate
that development followed an approved procedure without revealing the raw
pixels to the verifier, but constructing efficient proofs for neural and
image-processing pipelines remains computationally expensive and
circuit-intensive. Split or secure multiparty computation could
prevent any single cloud provider from receiving the complete plaintext image,
although these approaches would introduce substantial computation and latency
for large raw photographs. Finally, a user-selected confidential-cloud design
could retain hardware-backed trusted execution while allowing the user to
choose a provider or jurisdiction. This would reduce centralization, but would
move, rather than eliminate, trust, and would introduce interoperability and key
management challenges. These alternatives demonstrate that stronger data
minimization is possible, but each requires a tradeoff between provenance
assurance, privacy, computational cost, operational complexity, and trust
distribution.
\par
Based on our Python testbed results, we recommend that ARI and PCC protections
go beyond verifying a workload's identity. Our experiments show that two
authorized workloads can pass the same attestation and transparency-log checks
and return identical photographs even when one secretly retains a gradient that
reconstructs the processed image. PCC should therefore enforce a simple
non-learning policy that prohibits the ARI workload from computing or retaining
gradients, embeddings, model updates, or other unnecessary image-derived state.
The workload should have access only to temporary memory required for image
verification and development, with persistent storage and outbound
communication disabled unless explicitly required and independently audited.
Apple should document exactly which data enters PCC, which derived information
may leave it, and what information is retained after processing. Finally,
independent reviewers should verify the complete ARI workload and its storage
and communication permissions, because attestation, transparency-log inclusion,
identical outputs, and small timing differences do not by themselves demonstrate
that the workload performs only the intended image-development operation.

\section{Conclusion}

PCC strengthens ARI provenance precisely because it receives the signed digital
negative and becomes responsible for validating and developing it. That same
design creates a privileged machine-learning access point to authenticated raw
pixels. Encryption protects the DNG in transit; it does not hide the image from
the authorized workload that must process it.

Remote attestation and transparency provide valuable evidence about which
approved release handled a request. They do not, without semantic policy or
verification, prove that the workload did not compute a gradient, embedding, or
model update. The formal counterexample shows that both the baseline and learning
workloads can be authorized, return identical results, and pass the same
identity-based acceptance policy, even though only one changes the persistent model
state.

This work does not claim that Apple currently uses ARI data for model training.
Instead, it argues that the public assurance that ``even Apple cannot access
the data'' should be interpreted carefully. PCC may effectively prevent routine
human access and access by unauthorized infrastructure, but the behavior of
Apple-authorized workloads remains a semantic trust assumption. Closing this
assurance gap requires independently verifiable guarantees that authorized
workloads cannot learn from or retain information derived from ARI photographs;
verifying workload identity alone is insufficient.

\balance

\end{document}